\documentclass[conference]{IEEEtran}
\IEEEoverridecommandlockouts
\usepackage{cite}
\usepackage{amsmath,amssymb,amsfonts}
\usepackage{algorithmic}
\usepackage{graphicx}
\usepackage{textcomp}
\usepackage{xcolor}
\def\BibTeX{{\rm B\kern-.05em{\sc i\kern-.025em b}\kern-.08em
    T\kern-.1667em\lower.7ex\hbox{E}\kern-.125emX}}
\begin{document}


\title{Towards Decentralized Identity Management in Multi-stakeholder 6G Networks\\
{\footnotesize }
}

\author{\IEEEauthorblockN{Sandro Rodriguez Garzon}
	\IEEEauthorblockA{\textit{Service-centric Networking} \\
		\textit{Technische Universit\"at Berlin / T-Labs}\\
		Berlin, Germany \\
		sandro.rodriguezgarzon@tu-berlin.de}
	\and
	\IEEEauthorblockN{Hakan Yildiz}
	\IEEEauthorblockA{\textit{Service-centric Networking} \\
		\textit{Technische Universit\"at Berlin / T-Labs}\\
		Berlin, Germany \\
		hakan.yildiz@tu-berlin.de}
	\and
	\IEEEauthorblockN{Axel K\"upper}
	\IEEEauthorblockA{\textit{Service-centric Networking} \\
		\textit{Technische Universit\"at Berlin / T-Labs}\\
		Berlin, Germany \\
		axel.kuepper@tu-berlin.de}
}

\newcommand{\ts}{\textsuperscript}

\maketitle

\begin{abstract}

Trust-building mechanisms among network entities of different administrative domains will gain significant importance in 6G because a future mobile network will be operated cooperatively by a variety of different stakeholders rather than by a single mobile network operator. The use of trusted third party issued certificates for initial trust establishment in multi-stakeholder 6G networks is only advisable to a limited extent, as trusted third parties not only represent single point of failures or attacks, but they also cannot guarantee global independence due to national legislation and regulatory or political influence. This article proposes to decentralize identity management in 6G networks to enable secure mutual authentication between network entities of different trust domains without relying on a trusted third party and to empower network entities with the ability to shape and strengthen cross-domain trust relationships by the exchange of verifiable credentials. A reference model for decentralized identity management in 6G is given as an initial guide for the fundamental design of a common identity management system whose operation and governance are distributed equally across multiple trust domains of interconnected and multi-stakeholder 6G ecosystems.  

\end{abstract}


\section{Introduction}

Over the past years, mobile network operators (MNOs) have gradually started to open their public land mobile networks (PLMNs) to 3\ts{rd}-party technology operators by partly outsourcing the operation and maintenance of even critical network functionality. This trend is not exclusively rooted in the technical supremacy of 3\ts{rd}-party solutions but is also the result of economical rationales or regulatory enforcement. Consequently, 6G networks become multi-tenancy systems \cite{An.2021} in which connectivity is considered a multi-domain service that relies on resources distributed across several trust domains \cite{Bertin.2021}. Despite the many advantages of orchestrating a group of independent technology operators to provide seamless and economically viable connectivity under the roof of a single MNO, it comes along with the challenge to establish trust between network entities of different administrative domains. 

A key role in trust establishment within trustless environments is played by the authenticity check of a network entity's identity. Even though certificate-based authentication with TLS 1.2/1.3 and X.509 certificates is today generally considered the preferred solution for authenticity checks in 5G, it is based on the key assumptions that there exists a common and independent trusted 3\ts{rd} party (TTP) in form of a certificate authority (CA) and that it securely and reliably keeps care of the validity of the network entities digital identities. But besides the severe risk of dealing with a single point of failure and attack, as of today, there exists no commonly trusted CA on a global level for roaming purposes in 4G/5G. In fact, the GSMA even recommends to operate at least one CA at each PLMN \cite{GSMAssociation.2021}, which then requires additional trust-based cross-certification between CAs of different trust domains \cite{EuropeanTelecommunicationsStandardsInstitute.2020}. In 6G, the lack of common trust anchors turns even into an insurmountable obstacle for trustworthiness since trust domains in form of cross-domain network slices will span PLMNs that will be under the jurisdiction of different and potentially competing MNOs. Nevertheless, openness and trustworthiness are two of the proclaimed key design principles for a multi-player ecosystem architecture in 6G \cite{Tong.2021}. 

In 6G, certificate-based authentication mechanisms can only be deployed in a trustful manner if the required trust anchor is neither located in one of the trust domains of the communication entities involved, nor in one of its own since the independence of an institution can never be guaranteed due to geopolitical reasons. The latter requirement deems a trust model that enables trust through a TTP such as within the Web of Trust practically unsuitable for 6G. The only remaining option is to outline a cross-domain identity management (IDM) concept where the governance is equally distributed across all involved trust domains. A pivotal aspect of decentralizing the control and operation of an IDM system, without any single network entity or trust domain being in exclusive charge, is to form a commonly governed and therefore trusted repository for elements that are needed to establish trust. In the spirit of a decentralized public key infrastructure (PKI), it could be considered as a common single source of truth for identity-related public keys in 6G that are shared in a tamper-proof manner among all trust domains and used for cross-domain trust building mechanisms such as certificate-based authentication or, in more general, the attestation and verification of all kinds of credentials.    

This article explores the opportunities of decentralized IDM for 6G by identifying trust domains in multi-stakeholder 6G networks and potential touch points in between. The concept of Decentralized Identifiers (DID) \cite{WorldWideWebConsortium.822021}, as recommended by the W3C, and the closely related user-centric paradigm of Self-sovereign Identity (SSI) \cite{Toth.2019} are briefly introduced as standardized means to implement a commonly governed and trusted IDM system for 6G that complies with privacy-by-design principles. To take full advantage of the trust-building mechanisms enabled through the decentralized governance of digital identities, a reference model is presented in form of a conceptual blueprint for future implementations of a decentralized IDM system in 6G. 

Chapter {\ref{domains}} starts with the identification of trust domains and cross-domain touch points in 6G and is followed in Chapter {\ref{related}} with a discussion of recent approaches to decentralize IDM in cellular networks to mainly improve network access. Chapter \ref{DID_SSI} gives a brief introduction to the concept of decentralized identifiers and verifiable credentials while Chapter \ref{architecture} presents the DID-based reference model for a common IDM system in multi-stakeholder 6G networks. Chapter \ref{conclusion} summarizes the findings and discusses open issues.

\section{Trust in Mobile Networks}
\label{domains}
A {\bf{trust domain}} is defined to be an enclosed space comprising uniquely identifiable network entities which mutually trust each other. Trust is modeled as a {\bf{trust relationship}} in which a network entity trusts another network entity with respect to a subject of matter and to a certain extent. To form a trust domain, network entities must at least share a common understanding of the subject of matter and intensity of inherent trust applied between each other. Network entities may or may not have trust relationships with network entities of other trust domains. A {\bf{trust model}} is then defined to be a constellation of identifiable network entities of the same or different trust domains with their pairwise trust relationships. Network entities of the same or different trust domains must still verify the authenticity of each other's identity by the exchange of unique identifiers and proof artifacts in case they want to interact via a trustless communication medium.         

The PLMN's trust model with respect to home network access is rather simple with only two trust domains: subscriber and MNO. The trust relationship in between entities of both trust domains is based on the outcome of a know-your-customer process at the MNO which usually leads the subscriber to become a known trusted customer. However, the air interface is a trustless communication medium. To gain access to services offered by the MNO, a mobile subscriber needs to proof its identity towards the MNO by means of an international mobile subscriber identity (IMSI) and a pre-shared symmetric key hidden within the subscriber identity module (SIM). If the mobile subscriber is successfully authenticated towards the MNO and vice versa, there are no further trust establishing measures required since the MNO is aware of the services a customer is permitted to use.    

With the introduction of roaming, at least one additional trust domain is involved in the provision of mobile connectivity because a visiting MNO (vMNO) takes over the role of the connectivity service provider on behalf of the home MNO (hMNO). Authentication is thereby accomplished by providing the vMNO limited access to the hMNO's authentication center. This requires a trust relationship, which is legally protected by bilateral roaming agreements. While inter-PLMN signaling for roaming purposes was originally assumed to be trustful due to the SS7 network being considered a member-only trusted domain, today, inter-PLMN signaling is preferably conducted via a Security Edge Protection Proxy (SEPP) in 5G with TLS and pre-shared X.509 certificates for mutual authentication purposes. If Internetwork Packet Exchange (IPX) providers are in between, mutual authentication with TLS/X.509 to and from IPX providers is recommended as well. The trust model for roaming makes use of a TTP in form of the hMNO to establish a trust relationship between a visiting subscriber and the vMNO. Network service authorizations are provided by the hMNO besides authentication artifacts to the vMNO to shape and strengthen the temporary trust relationship.     

Until recently, a PLMN was considered a single trust domain because each critical network component and the links in between were operated by a single MNO. But the technological trend towards network function (NF) decomposition and radio access network (RAN) disaggregation breaks up this enclosed trusted environment and introduces new trust domains along the functional layer since each network component can be operated by an independent party within a dedicated trust domain. At least with the introduction of network slicing and the resulting virtual networks, individual NF instances must be assigned to at least one trust domain in the service-based architecture (SBA) and secured against unauthorized access by means of authentication and authorization. Consequently, the core network (CN) with its SBA in 5G evolved into a zero trust architecture (ZTA) in which mutual authentication between NFs with TLS and X.509 certificates was being introduced as an authentication option at the service-based interface (SBI). A similar approach with TLS, X.509 certificates and IPSec is discussed by the Open RAN alliance for the mutual authentication among the centralized (CU), distributed (DU), and radio units (RU) and the radio intelligent controller (RIC) of an open RAN.   

By decoupling the software of network components from the underlying physical hardware, the concept of NF virtualization enables network operators to let external cloud providers host NFs as pure software elements within their cloud or edge cloud infrastructures. This relieves MNOs from the burden to own or maintain physical hardware. However, the virtualization of execution environments introduces new trust domains up the system stack starting from the physical equipment provider, over the hypervisor manager and the orchestrator, till the software-defined networking controller and virtual NF operators. Trust building mechanisms between entities of the different layers will be required but are yet not standardized.                

MNOs also started to sell mobile towers for economic reasons and then lease them back from resource brokers as demand requires. This extends the scope of parties with their trust domains not only to independent hardware providers for the RAN, but likely also to resource brokers for both the access networks (ANs) and CNs in the future. With multi-access edge computing (MEC), MNOs will even open the RAN and its resources to application and service operators on the application layer to enable low-latency mobile applications and services. Besides the network exposure function (NEF) and service enabler architecture layer (SEAL) for verticals in 5G, this is another but yet to be defined touch point between entities of PLMN-centric and application layer trust domains. 

The presented list of expected parties, trust domains and the potential touch points in between is by no means a comprehensive but already an extensive snapshot of the trust problem space that is expected to emerge with the introduction of multi-stakeholder 6G networking. With each evolutionary step, more parties will likely be involved with their technical solutions or services in the operation and maintenance of a PLMN and the more trust domains and cross-domain touch points appear in the PLMN landscape. However, today's centrally managed, partly incompatible, and domain-specific identity and key management solutions have their limitations when it comes to trust building in trustless environments, as their usage requires a minimum level of trust towards the institution that runs the identity and key management system. The main objective is therefore to come up with a holistic and unified solution for identity and key management in 6G, which does not require mutual trust or trust in a 3\ts{rd} party to enable mutual authentication between 6G network entities of different trust domains, and which is, at the same time, capable to establish variously shaped trust relationships between these network entities for a wide range of 6G scenarios.

A common trust foundation among the network entities of all trust domains can be established by agreeing on a commonly accepted cross-domain IDM concept, similar to the global domain name system (DNS). But instead of letting non-government institutions in the role of TTPs such as the ICANN being in sole charge of IDM, the governance and operation of a global IDM system for 6G should be equally distributed among all trust domains. The decentralized operation should ensure the reliability, accessibility, and resilience of the global IDM system, while the decentralized governance enables trust relationships to be established between entities that are even located in geopolitically contrasting regions. So rather than having to trust a 3\ts{rd} party that is bound to national legislation, a globally operating IDM system itself should be seen as a technically enabled and entirely independent trust anchor.

\section{Related Work}
\label{related}

Despite the benefits of decentralized IDM being known since the appearance of the Web of Trust, its adoption in cellular network research was triggered by the successful application of {\bf{Distributed Ledger Technologies (DLTs)}} as a trust-enforcing technical foundation between entities of trustless peer-to-peer networks. At the intersection of cellular network and IDM research domains, distributed ledgers (DLs) are primarily interpreted as technical and commonly trusted means to persist and share data related to digital identities, e.g. public keys and certificates, in a secure, synchronized and tamper-proof manner among network entities of different trust domains. These identity-related proof artifacts within the DL are used to mutually authenticate network entities without involving a TTP and for other trust-building mechanisms such as the attestation and verification of credentials.    

To enable the registration of a massive amount of IoT devices in 5G networks, Jia et al. propose to manage IoT device identities via DLTs and to integrate DLT nodes into the edge network. Since the shared identities contain sufficient information to proof the authenticity of IoT devices, the authentication can only be executed at the edge node within the AN \cite{Jia.2020}. The distributed and trusted authentication system introduced by Guo et al. is based on similar ideas and comprises an additional DLT layer at the edge of the network to enable edge-only authentication of IoT devices \cite{Guo.2020}. By implementing IDM at the control plane in cognitive cellular networks with DLT, Raju et al. could proof that the signaling traffic related to network access can be reduced by up to 40\% \cite{Raju.2017}. In fact, within the problem domain of massive IoT device registration handling, DLT-based IDM solutions received attention as a promising approach to overcome latency issues and potential bottlenecks on the backhaul link \cite{Jia.2020, Fedrecheski.2020, Guo.2020, Raju.2017}. Fedrecheski et al. give a more general outlook of applying DLT-based IDM to the domain of the IoT by comparing it with the current certificate-based approaches \cite{Fedrecheski.2020}. The concept proposed in \cite{T.Hewa.2020} manages the validity of IoT device certificates with DLT to enable fog nodes in 5G networks to verify the device certificates securely during an attach request. Although not aiming at subscriber authentication, the user authentication scheme described in \cite{R.Almadhoun.2018} makes use of DLT core addresses to securely manage user access to IoT devices via DLT-connected fog nodes. Gnomon \cite{R.Ansey.2019} and the DIAM-IoT framework \cite{XinxinFan.2020} both apply the latest W3C proposed recommendation of the DID concept within their DLT-based IDM solutions for a TTP-less IoT device registration process \cite{XinxinFan.2020, R.Ansey.2019} and to securely obtain and verify software updates \cite{R.Ansey.2019}.        

Xu et. al. go beyond the domain of the IoT and describe a more general approach to enable subscriber authentication by means of DLT-based IDM \cite{Xu.2020}. Their approach completely decouples subscriber identification information from the data needed to authenticate towards a mobile network. The latter type of data is persisted in a DL, which enables subscribers to access mobile networks in a privacy-preserving manner without disclosing personal information. Haddad et al. follows a similar approach and introduces a new authentication and key agreement protocol for 5G that makes use of DLT-based IDM to enable 5G networks to authenticate visitors without querying the home PLMN authentication center \cite{Z.Haddad.2020}. With respect to the same objective, Yue et al. sketches the idea of a consortium DL to store the subscriber's public keys to facilitate the authentication of visiting subscribers \cite{Yue.2021b}. Yan et al. proposes to issue and manage X.509v3 certificates via a DL within mobile networks in order to be compatible with traditional centralized PKIs used today \cite{J.Yan.2020}. With this more generic and less intrusive approach of a decentralized PKI for mobile networks, Jan et al. can extend the use of DLT-based IDM to other actors within the PLMN ecosystem such as manufacturers, vendors, and service providers.    

The vast majority of approaches aim to improve the efficiency and to increase the privacy level of the subscriber's authentication procedure. With exception of the Gnomon \cite{R.Ansey.2019} and the DIAM-IoT framework \cite{XinxinFan.2020}, which rely on the W3C DIDs recommendation, they differ in the types and structures of the subscriber-related proof artifacts they persist in the DL. However, the potential of DLT-based IDM is not limited to subscriber-centric IDM on the control plane but can, as extensively discussed in \cite{RodriguezGarzon.2022} and briefly outlined in Chapter \ref{domains}, also be adopted to establish trust between entities of the application, data and management planes of 6G networks. For a holistic and unified DLT-based IDM solution for 6G that is suitable to handle identity-related proof artifacts for a wide range of known and yet unknown network entities, at least the structure of proof artifacts shared via a common DL needs to be unambiguously specified and standardized.

\section{Decentralized Identifiers} \label{dids}
\label{DID_SSI}

With the proposed recommendation of {\bf{Decentralized Identifiers (DIDs)}}, the W3C\footnote{World Wide Web Consortium} specified a domain-independent structure for elements of a digital identity that are meant to be persisted and shared in a tamper-proof manner among trust domains via a verifiable data registry (VDR). A DID is a unique identifier of a DID subject respectively a person or an entity of any kind and resolves to a corresponding DID document within a VDR. Trusted databases, decentralized databases, or DLs can serve as VDRs to store DID documents.  Only the DID subject or a deputy appointed by it is capable to modify the associated DID document. All DID subjects beyond have read-only access to other's DID documents. A DID document does not contain any personally identifiable information of the DID subject but associated verification material such as public keys. The verification material enables other DID subjects to verify the validity of personal identity information that can be shared, if desired, by the DID subject in a bilateral manner. The feature of making personal identity information verifiable becomes necessary since this privacy-sensitive information remains with the DID subject and is not meant to be provided by a TTP such as public authorities. Hence, a DID fulfills the fundamental requirement for a {\bf{Self-Sovereign Identity (SSI)}}, namely to pass the exclusive control over the digital identity to the subject represented by it. The technical representation of an entity's digital identity in the form of a uniquely resolvable DID document within a commonly trusted VDR enables TTP-less mutual authentication and TTP-based attestation and verification of credentials among entities of any kind. 

TTP-less mutual authentication empowers entities of the same or different trust domains that are represented by DID documents to autonomously set up secure and trustful end-to-end communication channels without involving any TTP. This is accomplished by storing a DID subject's public key for authentication purposes within the DID document. The associated private key remains at the premises of the DID subject. During an end-to-end connection establishment, both DID subjects make use of their own private and the other DID subject's public key within the DID documents in the DL to mutually verify the authenticity of each other's identity. The W3C did not specify a particular authentication method to apply which gives the DID subjects the freedom to choose a preferred one. Optionally, but as well not in the scope of the DID specification, symmetric session keys can be agreed upon to enable confidentiality through encryption. In principle, TTP-less mutual authentication with DIDs can be beneficially used anywhere in 6G where network entities of different trust domains are faced to trustfully interact with each other within a trustless environment. This includes NF-to-NF, MNO-to-MNO, AN-to-CN or UE-to-MNO communication realms. DID-based communication protocols have already been specified with the message-based, asynchronous and simplex DIDComm \cite{DecentralizedIdentityFoundation.24.01.2022} by the Digital Identity Foundation (DIF) being the one with a high maturity level.       

\begin{figure}[tp]
	\centerline{\includegraphics[scale=0.345]{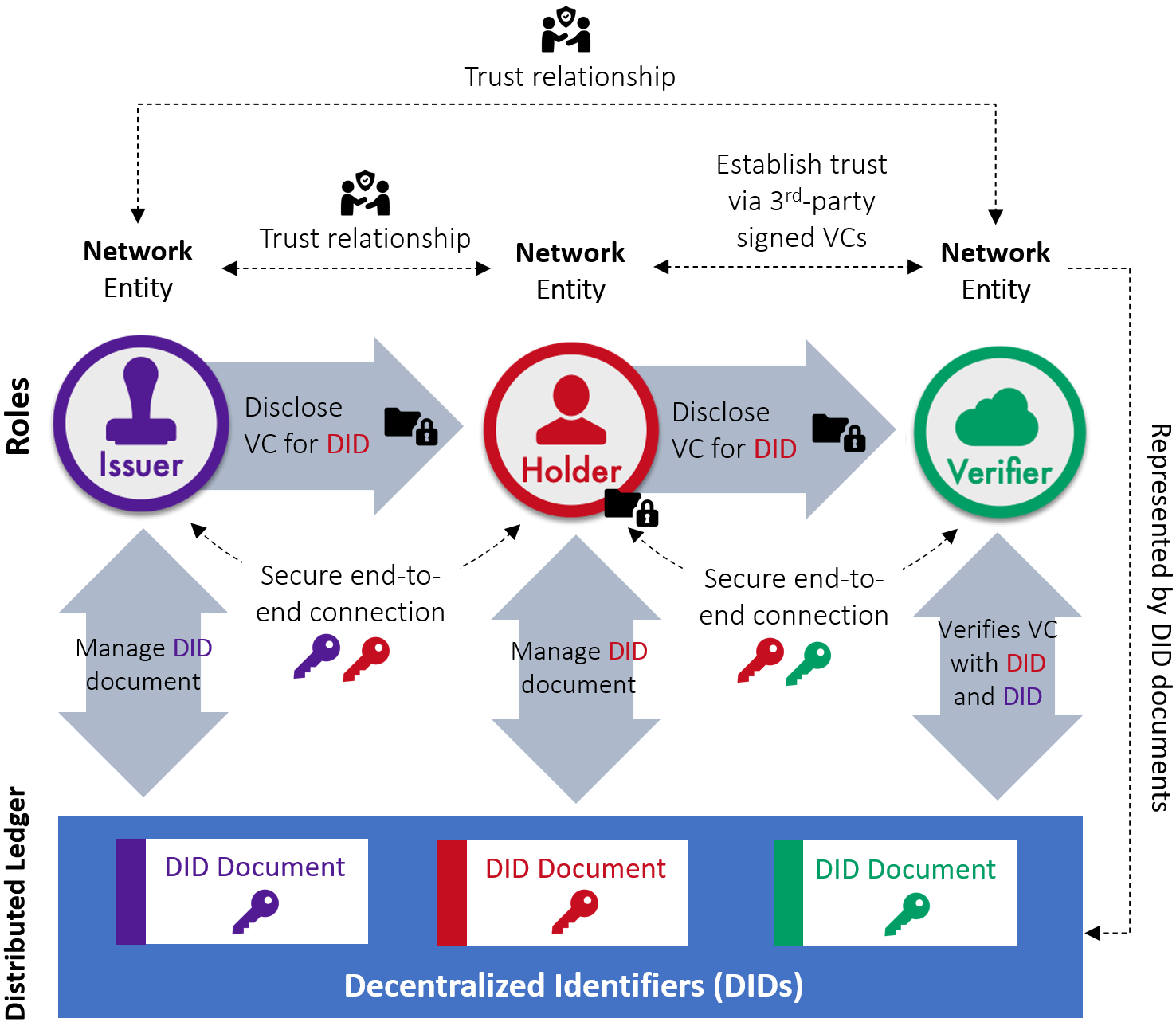}}
	\caption{DID-based attestation, verification and exchange of credentials.}
	\label{fig_SSI}
\end{figure}

Once established, the end-to-end communication channel can be used to securely exchange {\bf{verifiable credentials (VCs)}} \cite{WorldWideWebConsortium.05.11.2021} as a means to shape and strengthen the trust relationship between DID subjects. VCs are cryptographically verifiable information associated with the digital identity of a DID subject. The credibility of VCs for trust building between DID subjects is based on the principal assumption that there exists a 3\ts{rd} DID subject in the role of a TTP that attests the validity of the credentials, that are being shared among the DID subjects, by digitally signing them. The digital signature attached to the credentials of a DID subject can be verified by another DID subject by means of the TTP's public key within the TTP's DID document. The TTP takes hereby the role of an {\bf{issuer}} of VCs. The DID subject of which the credentials are being digitally signed by the issuer takes the role of the {\bf{holder}} of VCs and a DID subject verifying the holder's VCs for trust building purposes is called the {\bf{verifier}}. The cryptographic verifiability of a VC does not only guarantee that the credential has not been tampered with by a holder, but also that it represents the truth from the perspective the issuer. Just as with the transfer of VCs from the holder to the verifier, the transfer of the VCs from the issuer to the holder can be done over a DID-enabled end-to-end communication channel. A schematic overview of DID-based attestation, verification and exchange of credentials is given in Figure \ref{fig_SSI}. 

Potential areas where VCs can provide added value for MNOs, their customers, non-3GPP network operators and other parties operating in or interacting with the PLMN ecosystem are manifold \cite{RodriguezGarzon.2022}. So far investigated is to use VCs to encode network access permissions issued by the MNOs, hold by the subscribers and be verified during network attachment by hMNOs or vMNOs. However, VCs can be used to encode network access permissions for all kinds of network entities. For example, VCs can replace the JSON Web Tokens applied in OAuth 2.0 by the network resource function (NRF) to govern the access of NFs in the SBA. They can also encode hMNO-issued permissions to serve subscribers as guests for vMNOs or operators of non-3GPP ANs which can be verified by the subscribers. The release and deploy management of virtualized network components or underlying execution environments can benefit from VCs by improving the trust relationship between operators and software providers by means of manufacturer-issued VCs. 

\section{Reference Model for 6G}
\label{architecture}
The potentials of DIDs and VCs to change the way all sorts of network entities are identified, authenticated, and authorized in multi-stakeholder 6G networks are extensive. However, the decentralization of IDM with DIDs in distributed cellular networks requires MNO's to fundamentally rethink network governance. Even though MNOs will still take the leading role in the operation of future PLMNs, they will have to give up their claim to be exclusive administrators of network-wide digital identities in exchange for the benefits of trusted cross-domain communication. The rights to set up, modify or drop digital identities of MNO-owned and operated network entities whenever needed remain in place for the MNOs, but other PLMN actors and stakeholders in the spirit of self-sovereignty are granted the same rights for the digital identities of their network entities. While any actor involved in decentralized IDM can inspect all digital identities, the governance driving the underlying VDR makes sure no one has the exclusive right to prevent the setup, modification, and removal of others' digital identities.

\begin{figure}[tp]
	\centerline{\includegraphics[scale=0.39]{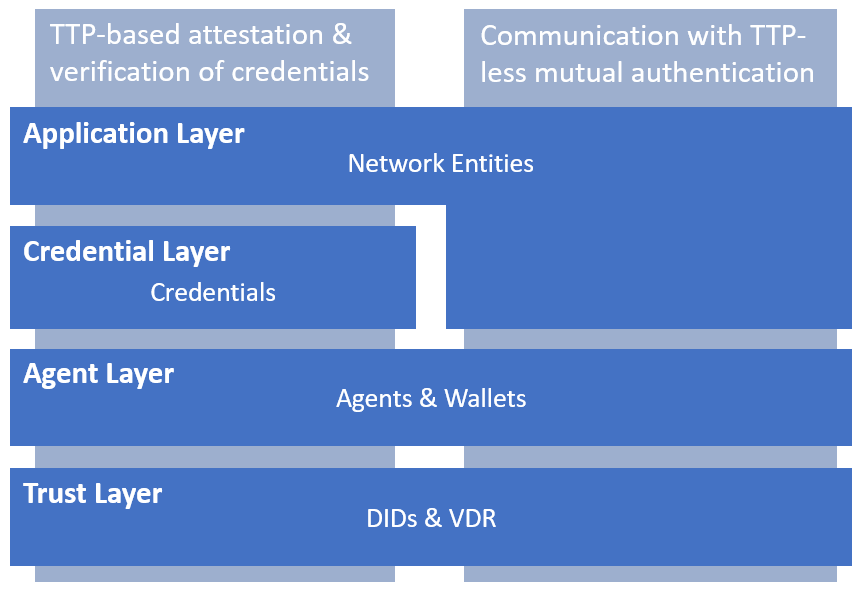}}
	\caption{6G Reference Model for Decentralized IDM}
	\label{fig_layers}
\end{figure} 

For the realization of decentralized IDM in 6G, a {\bf{reference model}} is needed that identifies the required components, groups them by responsibilities and clearly separates the task ranges from one another. It should consider only the two functionalities enabled by a decentralized IDM that are of particular interest for 6G, namely a) the communication with TTP-less mutual authentication and b) the TTP-based attestation and verification of credentials. Figure \ref{fig_layers} outlines the reference model for decentralized IDM in 6G. It borrows terminology from the SSI reference model specified by the Decentralized Identity Foundation (DIF) and extends it to cover the functionality of establishing communication channels with TTP-less mutual authentication.

At the {\bf{trust layer}}, VDRs form the technical foundation. They enable verification material of network entities to be shared in a tamper-proof manner in the form of DID documents across a trustless peer-to-peer network of 6G network entities that are associated with different trust domains. Agents in the sense of clients act in the {\bf{agent layer}} as the central interface between the network entities and the VDR, but also between the agents themselves. The credentials of the {\bf{credential layer}}, e.g. representing network access permissions, are thereby securely stored in software or hardware wallets. At the {\bf{application layer}}, network entities of all kind, ranging from a SIM of a subscriber's smartphone or IoT device, over a NF, cloud instance or a manufacture's firmware repository to an edge computing unit or single edge cloud application instance, make use of verification material provided by the agents to establish trust between each other.

\subsection{Trust Layer}
A VDR acts as a trust anchor between all network entities. It is intended to securely store the resolvable DID documents in a decentralized fashion. DLTs are the most common technical incarnation of VDRs. In 6G, a DL is envisioned to persist DID documents in a synchronized and tamper-proof manner among all network entities of interconnected trustless 6G PLMNs. Already persisted DID documents can't be removed, and new DID documents or changes to existing DID documents only be added. Since the DID document of each network entity is continuously replicated across all network nodes, all network entities have access to it but only the owner or a deputy is able to modify its DID document. 

Unlike a local TTP database for certificates, the shared state of the DL for DID documents must be agreed upon by the network nodes. Finding consensus on a common new state becomes necessary as soon as at least one network node creates a new or modifies an existing DID document within the DL. The way in which a valid state is agreed upon by means of a consensus algorithm is primarily derived from the desired network governance and the non-functional requirements that result from the use cases to be supported. Network governance specifies who can access the DL, add and modify DID documents, add or remove participants, and  who participates in the consensus finding process. Non-functional requirements resulting from the supported use cases comprise beside others scalability, security, energy consumption, signaling overhead and performance. The latter is measured in throughput and latency. Throughput is defined by the number of write transactions per second (TPS). The latency represents the time required to agree on the new state after someone added or modified a DID document. 


Each DLT supports different sets of network governance patterns and consensus algorithms and therefore exposes different and configuration-dependent non-functional properties. Which DLT and configuration to use in 6G and how to lay out a DL within a single or across PLMNs is therefore a matter of the desired network governance, the supported uses cases and the set of network entities targeted at the use cases. 

In a public permissionless DL for DIDs, all network entities can read and write DID documents without explicit permission. Finding consensus must be incentivized with rewards. A consensus can either be achieved by proof-of-work (PoW) or (delegated) proof-of-stake (PoS) algorithms. DLs applying PoW challenge participants to solve computational expensive artificial problems during the consensus finding process to fairly reward participants that contribute with computational power. This results in a high energy consumption, deeming PoW unsuitable for PLMNs. With PoS, participants own a share of the DL in form digital tokens. The more tokens a participant owns, the higher is the chance that the participant is selected to propose the new state. Hence, participants with high stake in the DL are incentivized to secure the status quo of the DL. DLs with PoS can achieve high TPS but the latency increases with the number of nodes. PoS-based DLs are prone to DDoS attacks because the consensus process can be halted by attacking at least 33\% of the nodes. By getting in possession of at least 66\% of the stake, the DL can be tampered with by a single actor. Unless every entity needs to publicly read and write DID documents, public permissionless DLs are of lesser practical use for the trust layer of decentralized IDM in 6G PLMNs.     



Public permissioned DLs for DIDs are suitable for 6G use cases in which all network entities can read DID documents but only a few are authorized to add or modify them. An example would be one for the control plane in which hMNOs and vMNOs manage their DID documents. Visitors are then able to verify VCs, provided by the vMNOs and signed by the hMNOs, that encode a hMNO's permission for vMNOs to serve their subscribers as guests. Reward-based incentives are not required since hMNOs and vMNOs as write-authorized participants have intrinsic motivation to maintain the DL in order to increase the efficiency of the subscriber registration process instead of earning rewards. The vulnerability to DL-halting or tampering attacks resembles approximately the one of PoS-based public permissionless DLs. Practical Byzantine Fault Tolerance (BFT) is mostly applied as the consensus algorithm. Despite high throughput, BFT suffers from a limited scalability. The synchronization overhead increases exponentially with every node, so does the related latency. BFT is therefore suitable for 6G use cases where the number of expected nodes remains manageable. However, governance rules can be adjusted by the participants through a consensus.    


In a private permissioned DL for DIDs, read and write operations require permissions. On-boarding new participants with read-only or read \& write permissions is strictly regulated by the network governance. BFT is mostly applied for consensus finding, with the same characteristics as in public permissioned DLs. However, private permissioned DLs take a step towards compliance with data protection regulations due to the strict access management. This type of DL qualifies for the majority of 6G use cases because access rules can be fine-grained and tailored to individual participants, the visibility of shared DID documents be limited to a defined set of participants, and the energy consumption results solely from the synchronization overhead and not from the computational power needed to solve computationally expensive artificial problems. For example, a private permissioned DL for DIDs of MNOs only would enable TTP-less mutual authentication on the control plane between MNOs without stressing the scalability constraints.

\subsection{Agent Layer}
\label{agent_layer}

Agents represent network entities. In the spirit of client software, they provide network entities transparent access to the components of the trust layer and are used to manage their DID documents and to obtain the other's DID documents. They can be located either directly or in the vicinity of the network entity, e.g., in the UE or a NF, or they can be operated as a cloud application. The agent's component that securely manages the own DID documents with the associated private and public keys is called the wallet. A pure-software wallet can exist within multiple agent instances for the same DID on different devices while other types of wallets can be bound to the hardware the agent is executing on. In 6G, a subscriber's wallet could, e.g., either be bound to a SIM or the SIM itself, as a trusted execution environment, could serve as the wallet. 

Agents can also be used by the network entities to establish secure direct communication channels between each other. This is accomplished, as outlined in Chapter \ref{DID_SSI}, with the help of the public keys within the network entities DID documents. The mutual authentication and the security measures to guarantee integrity and confidentiality of the communication channel are taken over by communication protocols, e.g., such as DIDComm. Agents sign, encrypt and decrypt thereby the messages exchanged by the network entities. The message format applied by DIDComm, e.g., is JSON Web Messages that basically specifies plain text envelopes with non-repudiable signatures. Since DIDComm ensures end-to-end security on the OSI application layer, it is agnostic to the transport protocol beneath. DIDComm messages can therefore be exchanged with any kind of transport protocol, e.g., with HTTP over TCP or even TLS. 

The verification material required to establish a DID-based communication channel between two agents do not necessarily need to be obtained from a DL. A DID document can also be self-certified and stored locally at the respective agent. To still enable mutual authentication by means of DIDs, the locally stored DID document needs to be transferred to the communication partner using out-of-band protocols. However, not involving the trust-enforcing DL reduces the level of trust that can be achieved. To nearly reach the same level of trust as with DID documents stored in the DL, VCs can be used among the agents after the communication channel is established to prove the authenticity of their identities through the attestation of a TTP. This reassembles the notion of X.509 certificates for authentication purposes but still allows the network entities to take advantage of other benefits of decentralized IDM with DIDs. For example, lightweight IoT devices without access to a DL that are yet unknown to an MNO and yet not represented through a DID document in a DL can make use of VCs to prove the authenticity of their identity towards the MNO by means of the corresponding IoT device operator's verification material in its DID document on the DL.

\subsection{Credential layer} 
A VC enables a network entity in the role of the holder to assert and prove parts of its digital identity to another network entity in the role of the verifier in order to extend their trust relationship. The reasons can be very diverse and include, among many others, authentication and authorization for service access. VCs may not but are preferable stored in the agent's wallet. A VC consists of credential metadata, a set of claims related to the holder's identity, and a proof. The metadata comprises a DID associated with the identity holder, the type of credential, and a timestamp of creation. The set of claims may include parts of the holder's identity, such as name or date of birth. The proof makes a VC tamper-resistant and cryptographically verifiable as it contains a signature from a key related to the DID of the issuer.

If a holder presents verifiable claims to a verifier, a verifiable presentation (VP) of a VC is created by the holder. A VP wraps the VC and signs it with the key related to the DID of within the credential's metadata. With the digital signature, the holder proves the ownership of the respective DID. This becomes necessary because a holder might not use the DID of the VCs metadata section but a locally stored and self-certified DID document to establish the communication channel with the verifier to prevent traceability. The additional proof also contains a random one-time nonce for a challenge between the holder and the verifier to avoid replay attacks. A VP is therefore meant for one-time use only.        

Although DID documents can be used for mutual authentication purposes, they are accessible to all participants of the DL and due to the way DLTs are implemented, they are not removable from the DL. This might violate national data protection regulations such as the EU General Data Protection Regulation (GDPR) because the pseudo-anonymous DID and the verification material in the DID document may be interpreted as personally identifiable information. The right for personal data confidentiality and the right to be forgotten may therefore not be compatible with the DID concept, if and only if DID documents are used in an inappropriate way to represent network entities for which data protection regulations such as the GDPR apply, namely EU citizens or, in more general, private persons. For network entities of businesses such as MNOs and data exchanged between businesses, bilaterally negotiated data protection rules apply instead of, e.g., the GDPR. 

In case of private persons as subscribers, a different approach needs to be taken to verify the authenticity of the private person's identity because she or he might not be represented by a DID document in a DL. Here, as an example, VCs can be used for subscriber authentication purposes towards, e.g., a vMNO while the mutual authentication with DIDs for communication establishment can be conducted separately with spontaneously created and self-certified DID documents, denoted as pairwise DIDs. In this scenario, a private person in the role of a subscriber is assumed to be identifiable by a DID but the corresponding DID document is only hold by the subscriber and is not persisted in a DL. During on-boarding, the hMNO (in the role of the issuer) hands out a VC to the subscriber stating that the identity referred by the subscriber's DID is indeed a valid customer of the hMNO. If the subscriber discloses the VC in form of a VP to a vMNO (in the role of the verifier) via a communication channel that was established using a pairwise and therefore entirely unrelated DID, then the subscriber needs to proof towards the vMNO that she or he owns the subscriber's DID. Otherwise, the vMNO can declare the received VP as valid with respect to the digital signature of the hMNO but it remains unclear for the vMNO whether the network entity that provided the VP is the one referred to by the subscriber's DID in the VP. Fortunately, as mentioned above, a VP contains the required additional ownership proof. 

\section{Compatibility with legacy systems}


With the VC data model \cite{WorldWideWebConsortium.05.11.2021}, the W3C proposed a generic JSON format to encode verifiable claims. Even though it is recommended to use VCs based on DIDs, it is not prescribed how the keys used to sign and verify the VCs are managed by the network entities. Decentralized IDM with public keys in DID documents and private keys at the network entities is just one implementation option. VCs could alternatively be issued by the strictly hierarchical PKIs found in MNOs today, provided to the holder in the spirit of X.509 certificates and could be verified by other network entities that share a common trust into the hierarchical PKI. So instead of binding a claim to a DID, it could be bound to any kind of identifier used in mobile networks today, e.g., an IP address. However, to enable a VC holder to proof ownership over the identifier, it should either take the form of a public key for which the holder owns the private key or it exists a different way for the verifier to authenticate the holder, e.g., via X.509 certificates. The particular advantages of decentralized IDM cannot be exploited without DIDs, but an important evolutionary step can be taken towards a uniform solution for the exchange of credentials, which can already be incrementally rolled out with legacy systems in PLMNs today. Since VCs are agnostic to the transport protocol, TLS in the SBA or envisioned TLS in the disaggregated RAN could be used instead of, e.g., DIDComm for the exchange of VCs. For example, VCs can then transparently replace the JSON Web Tokens of OAuth 2.0 that are used to grant access to NFs.

\section{Conclusion and Future Work}
\label{conclusion}

This article investigated the immense potential of decentralized IDM in multi-stakeholder 6G networks to enable mutual authentication between network entities of different trust domains without relying on a TTP, and to empower them with the ability to shape and strengthen their trust relationships through the exchange of VCs. Various trust domains have been identified whose network entities will need to rely on the respective mechanisms to establish and sustain trust with their counterparts of other administrative domains in order to provide mobile connectivity jointly in trustless interconnected 6G networks. A reference model for decentralized IDM in 6G was presented, which is intended to serve as an initial guide for the fundamental design of a common IDM system whose operation and governance is distributed equally across all trust domains of a 6G ecosystem. However, a number of issues still need to be resolved before decentralized IDM can be widely adopted in 6G and future converged networks. 

Even though an upcoming standard for the unique identification of entities and the decentralized management of associated verification material is realized with the introduction of DIDs, it remains open how DIDs are to be designed in the respective 6G context and how a decentralized IDM system supporting DIDs is implemented and seamlessly integrated into the 6G landscape. While the concrete use-case specific design of DIDs is a matter of standardization and the integration a matter of an actor-specific deployment configuration, the implementation requires a common agreement on the DLT to use for a single or a range of use cases and on the related governance rules to apply.  The practical suitability of state-of-the-art DLT for decentralized IDM in 6G with respect to critical non-functional requirements such as low energy-consumption, guaranteed update speed, scalability, low signaling overhead for synchronization and no susceptibility to attacks, which give single trust domains or malicious federations of a few trust domains temporal full control over distributed ledgers, needs to be further investigated and evaluated thoroughly. 

\bibliographystyle{IEEEtran}  
\bibliography{references}

\end{document}